\documentclass[a4paper, 12pt, oneside]{article}

\title{Active Learning Driven Materials Discovery for Low Thermal Conductivity Rare-Earth Pyrochlore for Thermal Barrier Coatings}

\author{A. Chowdhury$^\text{a}$, A. Rinc\'on Romero$^\text{a}$, G. Figueredo$^\text{b}$, T. Hussain$^\text{a}$}

\date{2024}

\usepackage{amssymb}
\usepackage{amsmath}
\usepackage{amsfonts}
\makeatletter

\usepackage{graphicx}

\usepackage{float}

\usepackage[hidelinks]{hyperref}

\usepackage{enumitem}

\usepackage{epigraph}

\usepackage{csquotes}

\usepackage[utf8]{inputenc}

\usepackage[english]{babel}

\usepackage{amsthm,thmtools}
\usepackage[nottoc]{tocbibind}

\usepackage{caption}
\usepackage{subcaption}

\usepackage{longtable}

\usepackage{booktabs}
\usepackage[table,xcdraw]{xcolor}
\usepackage{pdflscape}
\usepackage{rotating}

\usepackage{multirow}

\usepackage{physics}

\usepackage[mathscr]{euscript}

\usepackage[numbers,sort]{natbib}

\usepackage{listings}
\usepackage{hyphenat}

\newcommand*{\rom}[1]{\expandafter\@slowromancap\romannumeral #1@}

\usepackage[margin=1.2in]{geometry}

\allowdisplaybreaks

\graphicspath{{Figures/}}

\usepackage[type=none]{fgruler}

\begin{document}

\begin{titlepage}
   \begin{center}
       \vspace*{1cm}
       \Huge
       \par\textbf{\@title}

       \Large
       \vspace{0.5cm}
            
       \vspace{1.5cm}

       \textbf{\@author}

       \vspace{0.5cm}
       \begin{itemize}
           \item[\small$^\text{a}$] \small Centre of Excellence in Coatings and Surface Engineering, University of Nottingham, University Park, Nottingham, NG7 2RD, UK
           \item [\small$^\text{b}$] \small Faculty of Medicine and Health Sciences, , University of Nottingham, University Park, Nottingham, NG7 2RD, UK
       \end{itemize}

       \vfill
            
       \vspace{0.8cm}      
   \end{center}
\end{titlepage}
\newpage
\begin{abstract}

High-Entropy/multicomponent rare-earth oxides (HECs and MCCs) show promise as alternative materials for thermal barrier coatings (TBC) with the ability to tailor properties based on the combination of rare-earth elements present. By enabling the substitution of scarce or supply-risk rare-earths with more readily available alternatives while maintaining comparable material performance, HECs and MCCs offer a valuable path towards alternative TBC material design. However, navigating this search space of compositionally complex materials is both time and resource intensive. In this study, an active learning (AL) framework was employed to identify HEC/MCC materials with a pyrochlore structure, with acceptable thermal conductivity (TC) for TBC applications. The AL framework was applied through a Bayesian optimisation (BO) strategy, coupled with a random forest surrogate model. TC was selected as the optimisation criterion as that is the most basic requirement of TBC materials. Over two iterations of the AL cycle, four compositions were generated and synthesized in the lab for experimental evaluation. The first iteration yielded two single-phase pyrochlores, $(La_{0.29}Nd_{0.36}Gd_{0.36})_2Zr_2O_7$ and $(La_{0.333}Nd_{0.26}Gd_{0.15}Ho_{0.15}Yb_{0.111})_2Zr_2O_7$, with measured thermal conductivities of 2.03 and 1.90 $W/mK$, respectively. The surrogate model predicted a TC of 2.009 $W/mK$ for both compositions, demonstrating it's accuracy for completely new compositions. The second iteration compositions showed dual-phase when synthesized, highlighting the need to take into account phase formation in the AL framework.
\end{abstract}

Keywords: Machine learning; Bayesian Optimisation; Thermal conductivity; AI; Ceramic Coatings

\newpage

\section{Introduction}\label{Intro}

Nickel-based superalloys that make up gas-turbine engines, are limited to operating temperature of 850 $^oC$ \cite{lagow_gas_turbine_temp_2016, mouritz_superalloys_2012}. Thermal barrier coatings (TBC) are a thermally insulating coating applied on various parts of the gas-turbine engine, that allow higher operating temperatures. Since the 1970s, yttria-stabilized zirconia (YSZ) has been the industry standard TBC due to its thermal and mechanical properties \cite{Stecura1978EffectsOC,cao_vassen_stoever_2004,Vasen2022-oq}. YSZ allows for operating temperatures of upto 1200°C, above which the materials suffers from phase instability \cite{liu_HEC_2022}.  Increasing turbine temperatures improves efficiency and reduces $CO_2$ emissions, driving the need for improved thermal barrier coatings (TBCs) \cite{darolia_2013}. Furthermore, TBC life-span is greatly reduced due to erosion by calcium-magnesium-alumino-silicate (CMAS), essentially sand and volcanic ash. This drives the need to explore alternative materials with lower thermal conductivity and higher CMAS resistance \cite{CMAS_multicomp_Rare_Earth_Li2023-om}. However, replacing YSZ requires materials with lower thermal conductivity, superior phase stability, and compatibility with nickel-based superalloys \cite{bakan_vaßen_2017, mehboob_TBC_failure_mechanisms_2020}. 

Among the materials studied most extensively for thermal barrier coating (TBC) applications are pyrochlore-structured rare-earth oxides \cite{bakan_vaßen_2017, vassen_zirconates_TBC, fergus_zirconia_pyrochlore_TBC_2014, wu_wei_low_TC_RE_Zirconates}. These materials have improved thermal properties, compared to YSZ, and better resistance to reaction with CMAS \cite{vassen_zirconates_TBC, TC_zhu_meng_zhang_li_xu_reece_gao_2021}. Furthermore, various studies have shown that high-entropy and multi-component ceramic (HEC and MCC) pyrochlores show improved thermal and mechanical properties \cite{CWAN_LGZ, ren_LaYb_ReZr_HEC_2015, liu_thermal_prprt_LGZ_2014}. HECs and MCCs allow for tailored material properties by introducing multiple cation species in the $A$ and $B$ sublattices \cite{tsai_HEA_2014}. This provides a large search space for new TBC materials. Exploring this search space through traditional methods of experimental measurement and computational simulation is time-consuming. Experimental measurements are also costly in terms of equipment and consumables. Alternatively, designing new compounds \textit{in silico}, to screen candidates for further experiments, can  significantly accelerate research and development, and reduce the number of experimental variables (eg, microstructure, human error, etc.). First principle approaches based on density functional theory (DFT), molecular dynamics, Monte Carlo simulations are extensively used for \textit{in silico} screening of materials. However, they require significant computational resources and are mostly limited by theoretical foundations\cite{liu_Materials_Discovery_2017}.

Machine learning can significantly speed up the discovery process through near-instantaneous predictions of the desired properties. For instance, Zhang et al.\cite{Zhang2024_Pyrochlore_material_discovery_ML-dz} used a hybrid knowledge assisted data-driven DFT and experiments-based approach to design low thermal conductivity (TC), high-entropy pyrochlore oxides (HEO). 6188 $(5RE_{0.2})_2Zr_2O_7$ compositions were generated using a smart design approach through an explainable machine learning algorithm. Their TCs were predicted via machine learning and the thermal conductivity of the most promising candidate was verified through both first-principle calculations and experimental characterization. The study identified $(Sc_{0.2}Y_{0.2}La_{0.2}Ce_{0.2}Pr_{0.2})_2Zr_2O_7$ as a promising candidate for low TC, with the measured TC closely matching its predicted value.

A recurring issue for machine learning in the field of high-temperature ceramics is the limited amount of data available \cite{Small_data_materials_science_Xu2023-re}. Costs of experiments, in addition to a focus on the collection of small quantities of samples under controlled conditions tend to limit dataset sizes. This is further complicated by a lack of consistency in thermal conductivity values and experimental protocol of single-element rare-earth oxides across various literature \cite{CWAN_LGZ, Wright2019SizeDA}. This makes it difficult to train accurate and generalisable models to predict various material properties.

Active learning (AL) is a type of supervised machine learning approach that can help mitigate the challenges of data collection for cases where it is expensive or time-intensive. AL methods systematically identify the most informative samples to add to the training data by exploring input spaces under uncertainty\cite{BayesOpt_mat_disc_Kusne2020-qd, Pilania2021_Machine_learning_in_mat_sci}. This can reduce the number of experiments required to train a machine learning model to an acceptable degree of accuracy. For example, \textit{Sulley et al.} conducted a study on the efficient discovery of new high-entropy alloys (HEA) with stable phases using active learning\cite{AL_HEA_Sulley2024-mf}. Their AL trained model achieved a testing accuracy of $95\%$ with only $27\%$ of their dataset vs $94.6\%$ accuracy for a random forest model trained on $80\%$ of their full dataset of 2198 quinary HEAs. In general, AL relies on a surrogate model initially trained on available data to make predictions and measure uncertainty. An acquisition function is used to evaluate potential candidates for labeling based on the predicted values and prediction uncertainty provided by the surrogate model. The acquisition function subsequently guides the choice of the most informative, unexplored data points, so the system can learn more efficiently by focusing on where it's most uncertain \cite{lookman_active_learning_materials_discovery_2019}. 

AL has seen applications in materials discovery, primarily in pharmaceutical research for drug-discovery and recently for the discovery of new alloys and ceramics \cite{AL_mat_discStein2019-ck, lookman_active_learning_materials_discovery_2019}. In particular, it has seen uses for property optimisation of high-entropy ceramics which is of more interest for TBCs\cite{Zhang2024_Pyrochlore_material_discovery_ML-dz}. \textit{Leverant and Harvey} used AL to screen the entropy-forming ability (EFA) of 15,504 high-entropy carbides \cite{HEC_AL_mat_discvry_Leverant2024-rc}. The AL revealed 4 carbides with higher EFA values than those in the recorded literature. The AL framework was deemed generalisable to other material properties such as mechanical hardness and thermal conductivity and non-equimolar MCCs, although these were not included in the study. To the authors best knowledge, studies on the effectiveness of this method on non-equimolar, multi-component ceramics (MCC) does not exist. MCCs expand the search space for low thermal conductivity material and also show the potential for lower thermal conductivity than equimolar HECs. \textit{Wright et al.} studied 9 single-phase compositionally-complex flourite oxides (CCFO), including both high-entropy, equimolar oxides as well as medium-entropy non-equimolar oxides \cite{HEC_vs_CCC_Wright2020-fi}. Their study found that the non-equimolar medium-entropy CCFOs had lower thermal conductivity than the equimolar high-entropy compositions, presumably due to the clustering of oxygen vacancies in the lattice.

AL is generally applied in materials by using a Bayesian optimisation (BO) strategy  \cite{Park2023_mat_disc_AL_multi_obj_bayes_opt}. BO guides the search for new data-points with the aim of achieving a user-defined objective rather than just reducing uncertainty, allowing for the optimisation of new materials for specific properties \cite{Wang2021-pre-train_GP_for_Bayes_opt}. BO has been successfully applied in both materials processing and material discovery problems. \textit{Memon et al.} applied BO to optimise the processing parameters for deposition of a silicon coating via suspension plasma spray \cite{Active_learning_spray_Memon2024-qr}. The study used an initial dataset of 26 spray runs to train the surrogate model on predicting in-flight particle temperature and velocity of the silicon. An additional 6 data-points were added by the BO, resulting in a $52.9\%$ improvement in RMSE and a $8.5\%$ improvement in $R^2$. Furthermore, BO reduced the predictive uncertainty within the region of target variables associated with optimal coating quality, leading to more confident and accurate model predictions in the area of practical interest.

This study explores the use of active learning, using Bayesian optmisation with a Random Forest model, to optimise high-entropy and multi-component rare-earth pyrochlore oxides for low thermal conductivity.  By selecting an appropriate acquisition function, we test if the accuracy of the surrogate model can also be improved efficiently in parallel to searching for a low thermal conductivity composition. Furthermore, by using a Random Forest (RF) algorithm, we use the inherent feature importance provided by RF to study the features with the most relevant to designing low thermal conductivity multi-component pyrochlore oxides.

\section{}

\subsection{Material discovery}

\subsubsection{Active learning in materials discovery}

Active learning has been applied to materials discovery to design new materials that optimise a property of interest as well as to find the relationship between the design space and property space \cite{Material_disc_bayes_opt_Chitturi2024-ux}. In particular, Bayesian Optimisation has shown promise in this field due to it's ability to handle complex material-property relations and the ability to work with small datasets \cite{Jin2023_bayes_opt_materials_discovery_review-xg}.

Bayesian optimization (BO) is a strategy of active learning that forces the AL to search for data-points to satisfy a specific constraint and not just to reduce uncertainty. This is normally done via an optimization/acquisition function. The core idea behind BO is to use a surrogate model, typically Gaussian Process (GP) but Random Forest (RF) is also used, that provides a probabilistic representation of the objective, offering both mean predictions and uncertainty estimates, to approximate an acquisition function based on available observations \cite{Di_Fiore2024-AL_and_bayes_opt}. By optimizing this acquisition function, the BO process iteratively selects new evaluation points.

The selection of an appropriate acquisition function, for a given problem, is, thus, vital to guide the search for new materials, when applying BO. This leads to the dilemma of exploration vs exploitation \cite{Exploration_Exploitation_Bondu2010-ss}. Exploration is the labelling of data in non-sampled areas of the search space, to reduce local uncertainty and improve data distribution. This increases the diversity of the dataset, increasing the accuracy of the model over the entire search space. Exploitation identifies regions with potential optimal values of the target parameter previously sampled areas of the search space. The selection of an acquisition function determines how skewed the AL process is towards either exploration or exploitation. Common acquisition functions include Expected Improvement (EI), Probability of Improvement (PI), and Upper Confidence Bound (UCB). 

\begin{itemize}
    \item Probability Improvement (PI) measures the probability that the prediction of the surrogate model at a given location in the search space, is better than the best observation of the model so far. This is an exploitative strategy \cite{Di_Fiore2024-AL_and_bayes_opt}.
    \item Expected Improvement (EI) is widely used in Bayesian optimisation problems applied to materials science \cite{Di_Fiore2024-AL_and_bayes_opt, lookman_active_learning_materials_discovery_2019, Active_learning_spray_Memon2024-qr, Xue2016-qa_Accel_material_design_adaptive_design, Material_disc_bayes_opt_Chitturi2024-ux, Wei2025-cp_discv_lead_free_alloy_multi_bayes_opt}. This method takes into account the magnitude of improvement over the existing best data point, estimated for each new data point in the search space. EI will, thus, focus on regions of the search space where the magnitude of improvement over the current best observation is expected to be the highest. The EI value will increase both for data points where the predicted target value is higher than the current best target value, and for data points where the uncertainty is high. This allows for a balance between exploration and exploitation \cite{Di_Fiore2024-AL_and_bayes_opt}.
    \item Upper Confidence Bound (UCB) takes into account the predicted mean and the uncertainty at a new data point. By selecting points that have either high values predicted by the surrogate model, or where the prediction uncertainty is high, UCB offers a balanced approach between exploration and exploitation\cite{Shoyeb_Raihan2024-qa}.
\end{itemize}

A study by \textit{Farache et al.} used AL in conjunction with physics-based simulations to identify multiple principal component alloy (MPCAs) with high melting temperatures \cite{Farache2022-active_learning_high_melt_temp_alloy}. The study used random forest (RF) as a surrogate model to explore a search-space of 554 possible compositions, using different acquisition functions. The results demonstrated that acquisition functions that balance exploitation and exploration found the optimal composition with the least number of experiments.

A notable advantage of BO is its ability to incorporate prior knowledge and handle noisy observations, making it robust in practical applications. By intelligently selecting evaluation points, BO significantly reduces the number of function evaluations required to find the global optimum, thereby saving computational resources and time \cite{Wang2021-pre-train_GP_for_Bayes_opt}.

\subsection{Surrogate models}

In active learning, a surrogate model, generally trained on a small initial dataset, is used to evaluate new data points in the search space, providing predicted values of a property as well as the prediction uncertainty to evaluate the acquisition function. Gaussian Process (GP) and Random Forest (RF) are commonly used as surrogate models in active learning due to their inherent ability of both models to provide prediction uncertainty \cite{lookman_active_learning_materials_discovery_2019}. While GPs perform better than Random Forests on low-dimensional, numerical configuration spaces, Random Forests natively handle high-dimensional, partly discrete and conditional configuration spaces better where standard GPs do not work well \cite{Van_Hoof2021-ns_hyperparameter_optimiz_surrogate_models}.

\section{Methodology}

In this section we introduce the details of the methodology implemented for our case study. We describe the various steps of the active learning; the initial dataset selection and pre-processing, the construction of the surrogate model, the acquisition function and composition generation and finally the material synthesis and characterisation.

For this study, a Bayesian optimisation strategy is applied, using a Random Forest algorithm as the surrogate model. The surrogate model is trained on a small initial dataset from literature. This initial dataset contains 23 compositionally complex pyrochlore oxides with thermal conductivity data that is experimentally measured at room temperature. The surrogate model is used to evaluate Expected Improvement (EI) of new data points in the search space. Parameters describing the crystal structure of each composition is used as the input features for the surrogate model. The active learning suggests a combination of these input features each iteration. A brute-force search is then used to generate compositions that are a close match to these parameters. Multiple compositions are generated for each set of parameters, and are then produced in the lab for physical evaluation of thermal conductivity. The data is updated in the training dataset and used to run the next iteration of active learning.

\subsection{Active learning}

\begin{figure}[H]
    \centering
    \includegraphics[width=1\textwidth]{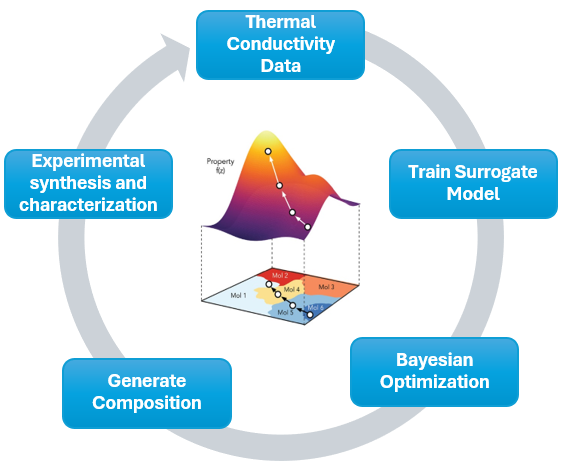}  
    \caption{Active learning workflow}
    \label{Img_AL_workflow}
\end{figure}

The active learning pipeline used in this work comprises of the following steps ; 1) Data gathering and pre-processing, 2) training the surrogate model to predict thermal conductivity of pyrochlore oxides, 3) define constraints for the Bayesian optimisation search space, 4) use surrogate model to evaluate thermal conductivity and uncertainty for a random initial set of compositions within the search space, 5) evaluate expected improvement (EI) of new data points using predicted thermal conductivity and model uncertainty, 6) Bayesian optimisation uses the EI to determine the next best spot in the search space, 7) the process is iterated \textit{n} times and the best input features that provide the highest EI is selected, 8) compositions are generated to match the target features and the two closest matches are selected, 9) compositions are synthesised and their thermal conductivity is measured, 10) finally the training database is updated with the new compositions and the measured TC values. A simplified version of this workflow is shown in Figure \ref{Img_AL_workflow}

\subsection{Thermal conductivity data}

The original data includes the thermal conductivity of 21 multi-component rare-earth oxide ceramics was gathered from the work of Wright et al. \cite{Wright2019SizeDA} to train the surrogate model. The data almost exclusively contains multi-component ceramic (mcc) compositions, with multiple elements in both $A$ and $B$ sublattices. This allows us to train a model that can account for these feature variations. The materials were synthesized as pellets from their binary oxides. Thermal diffusivity of the pellets was measured using LFA 467 \textit{HT Hyperflash} (NETZCSH, Germany), laser flash analysis. The specific heat  of each material was calculated using the Neumann-Kopp method\cite{suresh_TC_LZ_GZ, leitner_Neumann-Kopp_application_2010}. The specific heat and thermal diffusivity was then used to calculate the thermal conductivity. The same experimental process is used in this study to produce and characterise the compositions generated by active learning.

\begin{figure}[H]
    \centering
    \includegraphics[width=1\textwidth]{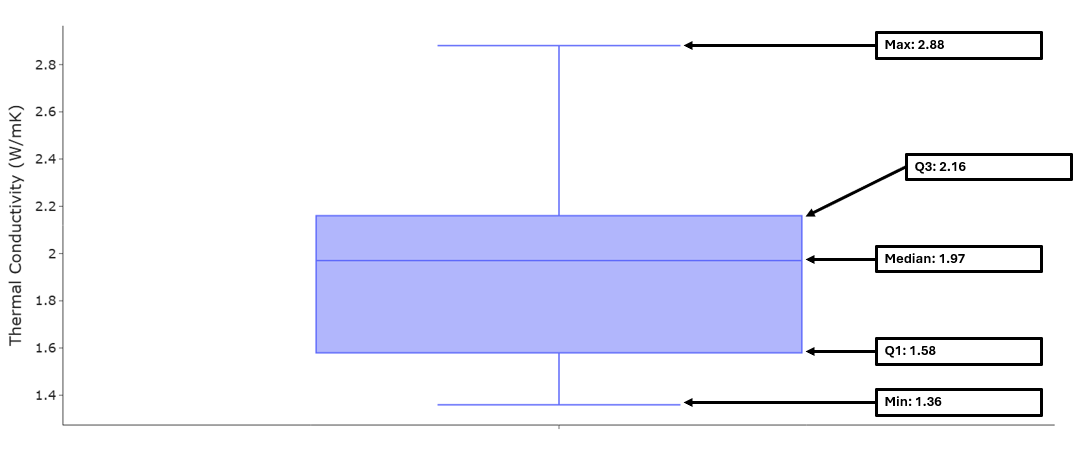}  
    \caption{Thermal conductivity distribution of training data}
    \label{Img_TC_dist_Wright_full}
\end{figure}

The distribution of thermal conductivity values measured from the compositions in this study are shown in Figure \ref{Img_TC_dist_Wright_full}. The composition with the highest TC of 2.88 $W/mK$ ($(Sm_{1/3}Eu_{1/3}Gd_{1/3})_2Ti_2O_7$) was treated as an outlier and removed from the training dataset. This was done to speed up exploration during the active learning towards regions of lower thermal conductivity as those are the regions of interest in this study. TC data is still sparse between the first and third interquartile ranges.

\subsection{Representation of pyrochlore structure}\label{Representation_of_pyrochlore_structure}

Prior to model training, the training data must be reformatted to capture relevant parameters (or descriptors) in a format suitable for machine learning. The descriptors used to train the surrogate model are listed in Table \ref{Table_model_features}. The primary parameters describe the $A$ and $B$ sublattice of pyrochlore compositions. Crystallographic descriptors were selected over compositional ones, as it proved difficult to implement a constraint on the search space to produce reasonable compositions. The package used for Bayesian optimisation only allow for a range to be specified for each descriptor and cannot ensure stoichiometry. Crystallographic descriptors correlating to thermal properties of pyrochlore oxides was studied by \textit{Yang et al.}\cite{yang_mechanical_thermal_properties_RE_pyrochlores_2018}. The study found that $R_A$, $R_B$, $M_A$ and $M_B$ have significant effect on thermal properties of pyrochlore oxides. The lattice configurational describes the distribution of elements in the $A$ and $B$ sub-lattices. This can be used to determine the total number of elements and atomic fraction of each element that should go into the $A$ and $B$ sub-lattices of the active learning compositions. Descriptor selection is described in more detail in a previous work \cite{Amiya_Data_representation_methods_2025-pl}.

\begin{table}[H]
\begin{center}
\begin{tabular}{cm{9cm}} 
 \hline
 Descriptors & Description\\
 \cmidrule(lr){1-1}\cmidrule(lr){2-2}
 $RA$ & Average ionic radii of $A$ cations\\
 \cmidrule(lr){2-2}
 $RB$ & Average  ionic radii of $B$ cations\\
 \cmidrule(lr){2-2}
 $MA$ & Average  atomic mass of $A$ cations\\
 \cmidrule(lr){2-2}
 $MB$ & Average  atomic mass of $B$ cations\\
 \cmidrule(lr){2-2}
 $Entropy$ & Lattice configurational Entropy\\
 \hline
\end{tabular}
\caption{ \label{Table_model_features}List of descriptors used for the surrogate model}
\end{center} 
\end{table}

The lattice configurational entropy is calculated via equation \ref{eqtn_dippo_entropy_metric} and \ref{eqtn_sublattice_entropy} from the work of Dippo et al. \cite{Dippo_config_entropy_2021}.

\begin{equation}
    EM = \frac{S_{SL/\text{mol atoms}}^{config}}{R}*L
    \label{eqtn_dippo_entropy_metric}
\end{equation}

Where $EM$ is the entropy metric, $R$ is the ideal gas constant, $L$ is the total number of sublattices and $S_{SL/mol_atoms}$ is the configurational entropy calculated using the sublattice model shown in Equation \ref{eqtn_sublattice_entropy}:

\begin{equation}
    S_{config}^{SL} = \frac{-R\sum_S\sum_ia^SX_i^S\ln{(X_i^S)}}{\sum_Sa^S}
    \label{eqtn_sublattice_entropy}
\end{equation}

Where $a^S$ is the number of sites on the $S$ sublattice and $X_i$ is the fraction of each individual species $i$ in the sublattice. For the $A_2B_2O_6O'$, $\sum a_S$ is 11 and $L$ is 5 due to the presence of the oxygen vacancy in the pyrochlore lattice.

\subsection{Surrogate model}\label{surrogate_model}

The surrogate model is trained to predict thermal conductivity. The model is trained using a Random Forest algorithm, which, as an ensemble method, yields both the mean prediction and its associated standard deviation. The standard deviation is used in place of model uncertainty, which, along with the mean prediction are required for the acquisition function to score the new data-point. Hyper-parameters were selected using the \textit{scikit-learn} implementation of the \textit{GridSearchCV} function. The parameters selected for optimisation are '\textit{n\_estimators}' and '\textit{min\_samples\_leaf}', which were set to ranges of 10 to 1000 and 1 to 5, respectively. Cross-validation (\textit{cv}) was set to 5 and number of jobs (\textit{n\_jobs}) to 1. The optimised hyper-parameters are listed in Table \ref{Table_model_1_hyperparams}.

\begin{table}[H]
\begin{center}
  \begin{tabular}{lc}
    \hline
    \ {Hyper-parameter} & {Constraints}\\
    \hline
    min\_sample\_leaf & 1\\
    n\_estimators & 10\\
    random\_state & 42\\
    \hline
  \end{tabular}
  \caption{Hyper-parameters for surrogate model used in first iteration of active learning}
  \label{Table_model_1_hyperparams}
\end{center}
\end{table}
At each iteration the surrogate mean absolute error (MAE) and $R^2$ of the model was evaluated using leave-one-out cross validation. Feature importances are evaluated using the \textit{shap} package \cite{shap_Lundberg2017-rg}. Feature importances are talked about in more detail in section \ref{feature_important}.

\subsection{Bayesian optimization and acquisition function}

Table \ref{Table_AL_stage_1_constraints} shows the pbound values (the upper and lower bounds of each input feature) used for active learning. The range of  $R_A$ and $M_A$ is set between ionic radii/atomic mass of lanthanum and ytterbium. $R_B$  and $M_B$ is set to the ionic radii/atomic mass of zirconium. The max entropy is set slightly above highest value in \textit{Wright et al.} database for a 10 element composition (2.93) (the composition is $(La_{1/7}Ce_{1/7}Pr_{1/7}Nd_{1/7}Sm_{1/7}Eu_{1/7}Gd_{1/7})_2
(Sn_{1/3}Hf_{1/3}Zr_{1/3})_2O_7$). This was done as the entropy of non-equimolar compositions can vary even for the same number of elements. Although yttrium is present in the training database, its parameters were excluded from the active learning constraints. The mass of yttrium is significantly lower than lanthanum, which has the next lowest mass amongst the potential A cation species (88.90585 u vs 138.90547 u). The rare-earth elements have a steady, linear increase in atomic mass as you move along the periodic table from lanthanum to ytterbium. This can lead to impossible combinations of atomic mass and cationic radii leading to the decision to exclude it. 

\begin{table}[H]
\begin{center}
  \begin{tabular}{lcc}
    \hline
    \ {Feature} &
      \multicolumn{2}{c}{Constraints}\\
    \hline
     & Min & Max\\
    \cmidrule(lr){2-2}\cmidrule(lr){3-3}\\
    $R_A$ & 0.985 & 1.16\\
    $R_B$ & 0.72 & 0.72\\
    $M_A$ & 138.90547 & 173.04\\
    $M_B$ & 91.224 & 91.224\\
    $\Delta{S_{config}}$ & 0 & 3\\
    \hline
  \end{tabular}
  \caption{Constraints on search parameters for active learning process}
  \label{Table_AL_stage_1_constraints}
\end{center}
\end{table}

Expected Improvement (EI) was selected for the acquisition functions, as it struck a balance between exploration and exploitation \cite{Bayesopt_EI_Brochu2010-ho, Active_learning_spray_Memon2024-qr}. The expected improvement is calculated using equations \ref{eqtn_EI} and \ref{eqtn_EI_z}.

\begin{equation}
    EI(x) = 
    \begin{cases}
        (max(f(x)) - {\mu}(f(x)) - {\varepsilon}){\Phi}(z) - {\sigma}(f(x)){\phi}(z), & {\sigma}(f(x))>0\\
        0, & {\sigma}(f(x))\leq{0}
    \end{cases}
    \label{eqtn_EI}
\end{equation}

\begin{equation}
    z = \frac{(max(f(x)) - {\mu}(f(x)) - {\varepsilon})}{{\sigma}(f(x))}
    \label{eqtn_EI_z}
\end{equation}

Where, $x$ is the combination of input parameters, ${\mu}(f(x))$ is the predicted thermal conductivity for $x$ given by the surrogate model, ${\sigma}(f(x))$ is the prediction uncertainty for $x$, ${\Phi}(z)$ is the standard normal cumulative probability density, ${\phi}(z)$ is the standard normal probability density, and $max(f(x))$ is the maximum thermal conductivity predicted using the current training dataset. As this is a minimisation problem, ${\mu}(f(x))$ is subtracted from $max(f(x))$.

\subsection{Feature importance}\label{feature_important}

Machine learning studies in materials science generally explores model explainability as opposed to treating the model as a black box \cite{Explainable_ML_Mat_Sci_Zhong2022-kw}. Feature importances is a set of techniques for evaluating the  relative effect of each input feature, used in the machine learning model, on the target variable of the problem. This is generally represented as a score or a ranking, with a higher ranking indicating that a certain input has more weight on predicting the target variable compared to the remaining inputs.

For this study, the SHAP (SHapley Additive exPlanations) package is applied to rank the contribution of each feature to thermal conductivity. These contributions are evaluated by changing each feature and measuring the change on the model output. For a single prediction, SHAP quantifies how each feature increases or decreases the prediction. By aggregating SHAP values across the dataset, the features are ranked using their average contribution \cite{shap_Lundberg2017-rg}.

\subsection{Generate compositions}

The parameters generated by the Bayesian optimisation are then converted to a $A_{2}B_{2}O_{6}O^{'}$ composition. To achieve this, a number of random $A_{2}B_{2}O_{6}O^{'}$ compositions are generated, their crystallographic parameters calculated, and then evaluated against the target parameters suggested by the Bayesian optimization. The evaluation is done via calculating the Euclidean distance between the two sets of parameters.

\begin{equation}
    D = \sqrt{(F_{1,c} - F_{1,t})+(F_{2,c} - F_{2,t}) + ....}
    \label{eqtn_Euclidean_distance}
\end{equation}

Where $D$ is the Eucledian distance, $F_n$ is feature $n$, $c$ represents the features for the randomly generated composition and $t$ represents the target features. To simplify the material synthesis, a number of constraints are set on the compositions that can be generated. $B$ element is set to zirconia only. A constraint, $N_{A}$, is also put on the number of elements allowed in the $A$ sublattice. The maximum number of allowable elements is set at 5 due to practical considerations regarding production and rare-earth metal costs. High-entropy compositions definitionally require 5 or more equi-molar elements. By setting the maximum value of $N_{A}$ to 5, the system can still potentially generate low, medium, and high-entropy compositions as well as multi-component ceramics (non-equimolar).

\subsection{Material synthesis}\label{Material_synthesis}

The samples generated for this study were produced via solution combustion synthesis of rare-earth nitrates and zirconium acetate precursors. The nitrates and acetate were mixed in appropriate quantities with deionized water. Citric acid was added as a stabiliser to achieve finer particle size \cite{MA_GZ_sintering_citric_acid_2015}. The mixture was then calcined in air in a box furnace at 1000 $^{o}C$ for 2 hrs, using a heating/cooling rate of 10 $^{o}C/min$. The resulting precipitate is then ball milled in a planetary ball-mill(Retsch PM 100 Planetary Ball Mill, Germany). The process used 2.5 $mm$ zirconia balls, a zirconia jar, and was done at 350 $rpm$ for 1 hr. Every 10 minutes, the milling was paused for 5 minutes to avoid overheating. The milling was done using ethanol in a $1:1:1$ volume ratio. The powder was pressed at 200 MPa using an isostatic cold press after having been initially pressed at 5 MPa using an uniaxial press in a 12.7 mm pellet die. The resulting green body is then sintered by being heat-treated in air, in a bottom loading furnace (Carbolite, UK), at 1600 $^{o}C$ for 6 hrs, at a heating and cooling rate of 5 $^{o}C/min$. The parameters for the heat-treatment were tuned to allow multiple, different compositions to be sintered at once with the aim of forming sufficiently dense ($\geq 90 \%$) pellets with pyrochlore phase. The synthesis conditions for this study were designed around ensuring both pyrochlore phase and high density (low porosity) for gadolinium zirconate. Gadolinium zirconate has an $R_A/R_B$ of 1.4625, close to the limit described earlier. Rare-earth element to the right of gadolinium, in the lanthanide series, always forms fluorites. Gadolinium zirconate also sintered significantly worse than lanthanum zirconate. Earlier experiments, indicated that a high temperature (1600 $^oC$) and a low heating and cooling rate would ensure pyrochlore phase formation and an acceptable level of sintering in gadolinium zirconate. 

\subsection{XRD}

XRD analysis was performed using the Bruker D8 system (Bruker, UK). The system is equipped with a copper-sealed tube x-ray source producing $Cu$ $k\alpha$ radiation at a wavelength of 1.54 $\AA$ and has a $2\theta$ range of 1-150$^o$. The analysis was done in Bragg-Brentano mode, with a step size of 0.02$^o$ and a time step of 0.2 $s$.

\subsection{Density}

Density was measured using the Archimedes method using the Ohaus density determination kit (Ohaus, Germany). The sample is first weighed dry (in air) and then wet (submerged in distilled water). The density is then calculated using Equation \ref{eqtn_Ohaus_density}.

\begin{equation}
    \rho = \frac{A}{A - B}(\rho_O - \rho_L) + \rho_L
    \label{eqtn_Ohaus_density}
\end{equation}

Where, $\rho$ is the density of the sample, $A$ is the dry mass of the sample, $B$ is the wet mass of the sample, $\rho_O$ is the density of distilled water at room temperature and $\rho_L$ is the density of air (0.0012 $g/cm^3$). The density of distilled water at various temperatures is provided in the supplementary information.

To calculate porosity, theoretical maximum density was calculated using the lattice parameters, which were, in turn, calculated using Bragg's Law, from the position of the $(622)$ peak (Equations \ref{eqtn_Bragg_law_1} and \ref{eqtn_Bragg_law_2}).

\begin{equation}
    a = \frac{\lambda*\sqrt{N}}{\sin{(\theta)}}
    \label{eqtn_Bragg_law_1}
\end{equation}

\begin{equation}
    N = h^2 + K^2 + l^2
    \label{eqtn_Bragg_law_2}
\end{equation}

Where $a$ is the lattice parameter, $\lambda$ is the X-Ray wavelength (1.54 \r{A}), $\theta$ is the position of the $(622)$ peak and $h$, $k$ and $l$ are the peak indices. The theoretical density was then calculated using equation \ref{eqtn_bulk_density};

\begin{equation}
    \rho = \frac{nM}{Na^3}
    \label{eqtn_bulk_density}
\end{equation}

Where $n$ is the total number of atoms per unit cell, $M$ is the molar mass and $N$ is Avogadro's constant. 

\subsection{Thermal conductivity measurement and correction}

Thermal diffusivity was measured using the laser-flash method (NETZSCH LFA 467 HyperFlash, Germany). The pellet dimensions were entered in the Proteus LFA analysis software (NETZSCH LFA Analysis, Germany) to calculate the thermal diffusivity. Thermal conductivity was then calculated using equation \ref{eqtn_TD_to_TC}:

\begin{equation}
    k' = \alpha\cdot{}C_p\cdot{}\rho
    \label{eqtn_TD_to_TC}
\end{equation}

Where, $k'$ is the thermal conductivity of the sintered pellet, $\alpha$ is the thermal diffusivity, $C_p$ is the specific heat capacity of the sample and $\rho$ is the density of the sintered pellet \cite{CWAN_LGZ}. The specific heat was calculated from the binary oxides using the Neumann-Kopp method \cite{kubaschewski_alcock_evans_1979, suresh_TC_LZ_GZ,wu_wei_low_TC_RE_Zirconates,leitner_Neumann-Kopp_application_2010}. The rare-earth zirconate pyrochlores, can be formed upon a reaction of its binary oxide constituents as shown:

\begin{equation}
    RE_2O_3 + 2ZrO_2 \to RE_2Zr_2O_7
    \label{eqtn_oxide_reaction_REZr}
\end{equation}

For low-entropy compositions ($RE^1_{(2-x)}RE^2_xZr_2O_7$), where $RE^1$ and $RE^2$ are two different rare-earth elements occupying the $A$-site, the reaction can be modified to:

\begin{equation}
    (1-\frac{x}{2})RE^1_2O_3 + \frac{x}{2}RE^2_2O_3 + 2ZrO_2\to RE^1_{(2-x)}RE^2_xZr_2O_7
    \label{eqtn_l_entropy_oxide_reaction}
\end{equation}

The specific heat can be calculated from the specific heats of the binary oxide constituents of the composition using equation \ref{eqtn_Neumann-Kopp}:

\begin{equation}
    C_p = C_{p(RE^1_2O_3)}(1-\frac{x}{2}) + C_{p(RE^2_2O_3)}(\frac{x}{2}) + 2C_{p(ZrO_2)}
    \label{eqtn_Neumann-Kopp}
\end{equation}

The oxide specific heats were calculated using equation \ref{eqtn_oxide_Cp}, where $T$ is the reference temperature (room temperature) and $a$, $b$ and $c$ are experimental coefficients compiled by \textit{Kubaschewski et al.} \cite{kubaschewski_alcock_evans_1979}. $ZrO_2$ is assumed to be in $\alpha$ phase for the coefficients. 

\begin{equation}
    C_p = a + bT + cT^{-2}
    \label{eqtn_oxide_Cp}
\end{equation}

Finally, the thermal conductivity is corrected for porosity using equation \ref{eqtn_TC_correction};

\begin{equation}
    \frac{k'}{k} = 1-\frac{4}{3}\phi
    \label{eqtn_TC_correction}
\end{equation}

Where $k$ is the thermal conductivity of a fully dense sample and $\phi$ is the porosity\cite{CWAN_LGZ}. To account for defects in the pellet affecting measurement, two pellets were synthesized for each composition. In the event that both pellets had significant defects, the average TC of the two pellets would be used.

\section{Results and discussion}

\subsection{New compositions generated by active learning}

\begin{figure}[H]
    \centering
    \includegraphics[width=1\textwidth]{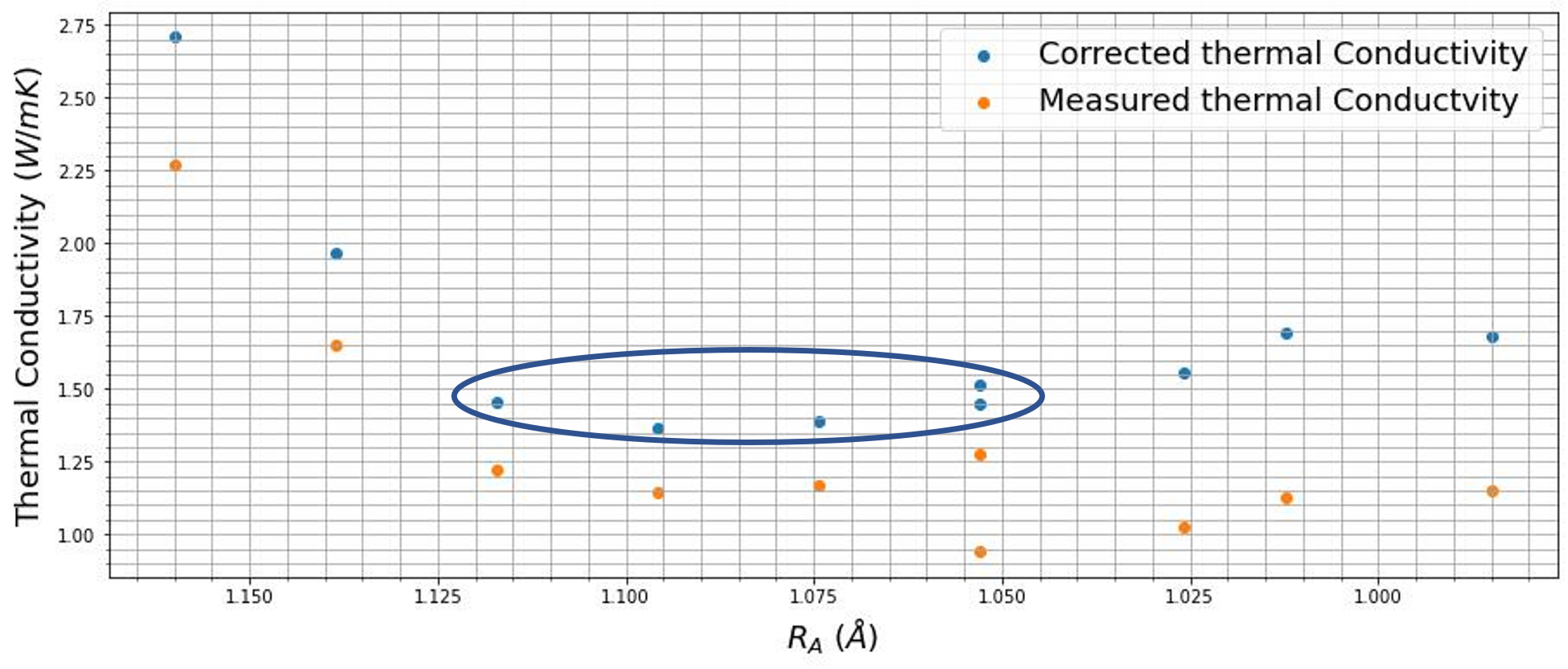}  
    \caption{Thermal conductivity vs $R_{A}$ of lanthanum gadolinium zirconate series}
    \label{Img_TC_LZ_GZ_series}
\end{figure}

The crystallographic parameters suggested by the AL are listed in Table \ref{Table_AL_new_cryst_params}, alongside their thermal conductivity predicted by the surrogate model. Figure \ref{Img_TC_LZ_GZ_series} shows the thermal conductivity (measured value and value corrected for porosity) of the lanthanum-gadolinium zirconate series, recreated from the work of \textit{Wan et al.}\cite{CWAN_LGZ}. The $R_{A}$ generated by the active learning is within the minimum TC range indicated in the figure. The $M_A$ of the compositions in this range is around 149 $u$, which matched up with the first iteration $M_A$. The $M_A$ suggested in the second iteration is significantly higher. 

\begin{table}[H]
\begin{center}
  \begin{tabular}{lccccccc}
    \hline
    \ {Iteration} & {$R_A$} & {$M_A$} & {$\Delta{S}_{config}$} & {EI} & {predicted TC}\\
    \hline
    1 & 1.1047 & 149.303 & 2.425 & 0.1216 & 2.009\\
    2 & 1.0898 & 171.3577 & 1.1236 & 0.0054 & 2.092\\
    \hline
  \end{tabular}
  \caption{Crystallographic parameters generated by active learning}
  \label{Table_AL_new_cryst_params}
\end{center}
\end{table}

The predicted thermal conductivity of the new compositions are also much higher than the minimum in the lanthanum gadolinium series. The median and Q3 of the TC in the training dataset are 1.97 and 2.145 $W/mK$, respectively. This would likely bias the predictions of the surrogate model towards this range of values.

\begin{table}[H]
\begin{center}
    \begin{tabular}{llllll}
        \hline
        Iteration          & Composition & RA & MA & Entropy & Distance \\
        \hline
        \multirow{2}{*}{1} & $(La_{0.29}Nd_{0.36}Gd_{0.36})_{2}Zr_{2}O_{7}$ &
                             1.1036 & 147.363  & 1.299 & 0.376        \\
                           & $(La_{0.333}Nd_{0.26}Gd_{0.15}Ho_{0.15}Yb_{0.111})_{2}Zr_{2}O_{7}$   & 1.0937 & 149.618 & 1.559 & 0.301        \\
        \hline
        \multirow{2}{*}{2} & $(La_{0.43}Nd_{0.05}Gd_{0.05}Dy_{0.09}Yb_{0.38})_{2}Zr_{2}O_{7}$                       & 1.0731 & 155.284 & 1.303 & \textunderscore \\
                           & $(La_{0.45}Nd_{0.05}Ho_{0.05}Dy_{0.1}Yb_{0.35})_{2}Zr_{2}O_{7}$     & 1.07565 & 154.78 & 1.314 & \textunderscore \\
        \hline
    \end{tabular}
\caption{Compositions matching generated parameters from active learning}
\label{Table_AL_cryst_params_to_compositions}
\end{center}
\end{table}

Table \ref{Table_AL_cryst_params_to_compositions} lists two compositions corresponding to each set of parameters. The compositions are not an exact match, and the configurational entropy, in particular, deviates significantly from the expected value. This is likely due to the constraints on the number of elements allowed in the $A$ and $B$ sublattices ($N_A$ and $N_B$), applied to the compositions. This was done to simplify production, but it may be worthwhile to remove this as a constraint for any future work.

\subsection{New composition characterisation}\label{New_comp_characterisation}

A total of 7 pellets were selected for testing. AL2-W-5-2 (no CIP) was created using only the uniaxial press, without the use of the isostatic press. Pellet identifiers as well as density and dimensions of each pellet are listed below.

\begin{table}[H]
\begin{center}
    \begin{tabular}{ll}
        \hline
        Composition          & Pellet ID\\
        \hline
        \multirow{1}{*}{$(La_{0.29}Nd_{0.36}Gd_{0.36})_{2}Zr_{2}O_{7}$} & 
                             AL-W-3-2\\

        \hline
        \multirow{1}{*}{$(La_{0.333}Nd_{0.26}Gd_{0.15}Ho_{0.15}Yb_{0.111})_{2}Zr_{2}O_{7}$} & AL-W-5-2 \\

        \hline
        \multirow{3}{*}{$(La_{0.43}Nd_{0.05}Gd_{0.05}Dy_{0.09}Yb_{0.38})_{2}Zr_{2}O_{7}$} & AL2-W-5-1(1)\\
        & AL2-W-5-1(2)\\
        & AL2-W-5 (no CIP)\\
        
        \hline
        \multirow{2}{*}{$(La_{0.45}Nd_{0.05}Ho_{0.05}Dy_{1.0}Yb_{0.35})_{2}Zr_{2}O_{7}$} & AL2-W-5-2(1)\\
        & AL2-W-5-2(2)\\
        
        \hline
    \end{tabular}
\caption{Pellet labels}
\label{Table_AL_comp_IDs}
\end{center}
\end{table}

\begin{table}[H]
\begin{center}
    \begin{tabular}{llllllll}
        \hline
        Pellet ID & Mean & Std. dev & Mean & Std. dev & Mean & Std. dev & Porosity\\
                  & thickness &     & diameter &      & Density & & ($\%$)\\
                  & (mm)      &     & (mm)     &      & ($g/cm^3$) & \\
        \hline
        AL-W-3-2 & 1.394 & 0.006 & 10.239 & 0.016 & 6.359 & 0.01 & \\
        \hline
        AL-W-5-2 & 1.403 & 0.001 & 10.266 & 0.022 & 6.489 & 0.002 & \\
        \hline
        AL2-W-5-1(1) & 1.833 & 0.013 & 10.354 & 0.037 & 6.471 & 0.002 & \\
        \hline
        AL2-W-5-1(2) & 1.79 & 0.006 & 10.309 & 0.018 & 6.698 & 0.002 & \\
        \hline
        AL2-W-5 (no CIP) & 1.591 & 0.042 & 10.344 & 0.057 & 6.474 & 0.001 & \\
        \hline
        AL2-W-5-2(1) & 1.803 & 0.018 & 10.154 & 0.0578 & 6.605 & 0.00 & \\
        \hline
        AL2-W-5-2(2) & 1.778 & 0.016 & 10.131 & 0.049 & 6.629 & 0.001 & \\
        \hline
    \end{tabular}
\caption{Pellet dimensions}
\label{Table_AL_pellet_dimensions}
\end{center}
\end{table}

\begin{figure}[H]
    \centering
    \includegraphics[width=1\textwidth]{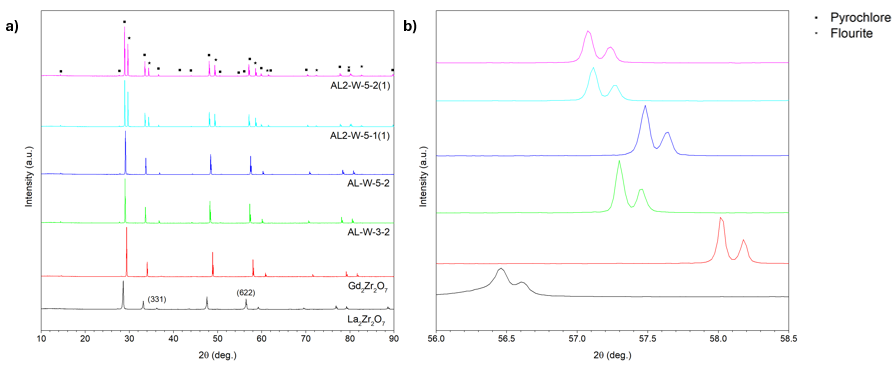}  
    \caption{XRD patterns of active learning generated compositions compared to LZO and GZO. The (622) peaks are shown magnified on the right.}
    \label{Img_AL-W_XRD}
\end{figure}

The XRD patterns (Figure (\ref{Img_AL-W_XRD})) show that the compositions generated in the first AL iteration (AL-W-5-2 and AL-W-3-2) are single phase pyrochlores. However, the compositions from the second AL (AL2-W-5-1 and AL2-W-5-2) iteration show both pyrochlore and fluourite phases. The $R_A/R_B$ of both compositions are 1.491 and 1.494, respectively. Typically, single-element rare-earth zirconates form the flourite phase when $R_A/R_B \leq 1.46$  \cite{fuentes_pyrochlore_ra_over_rb_criteria_2018}. However, synthesis conditions also affect the phase formation.

The formation of the pyrochlore phase may be complicated by the introduction of multiple cations within the $A$-site. Although work by \textit{Teng et al.} \cite{Teng2021-mz_HEC_phase_formation} and \textit{Liu et al.} \cite{Liu2023-ik_HEC_phase_formation} indicate that the ratio rule holds for HECs. However, the compositions in question, in this study, are MCCs, which is non-equimolar in contrast to HECs. It's also possible that this was error due to the very low amounts of some elements in the AL2 series ($Nd$, $Gd$ and $Ho = 0.05$). Further work will be needed to investigate this. However, the simpler solution may be to constrain the model to avoid atomic fractions $<0.1$ and perhaps even reduce the allowable $R_A/R_B$ range.

\subsection{Thermal diffusivity measurements}

\begin{itemize}
    \item Constraint on number of elements in A and B sublattice interfering with configurational entropy as a parameter? Better to remove entropy for main runs and have one run with entropy and no constraints on A and B to see generated composition?
\end{itemize}

\subsection{Thermal conductivity measurement of compositions}

Thermal diffusivity (TD) of each pellet is shown in table \ref{Table_AL_pellet_TD}. The TD of different compositions generated in the same active learning iteration are roughly similar. This is a good indication that the features selected to train the surrogate model are decent predictors for thermal conductivity. That said, it is difficult to make this conclusion due to the small sample size and especially considering the added complication of dual-phase in iteration 2.

\begin{table}[H]
\begin{center}
    \begin{tabular}{llll}
        \hline
        Iteration          & Composition & Thermal diffusivity & St. Dev\\
                           &             &      ($mm^2/s$)\\
        \hline
        \multirow{2}{*}{1} & AL-W-3-2 & 0.836 & 0.002\\
                           & AL-W-5-2 & 0.729 & 0.051\\
        \multirow{3}{*}{2} & AL2-W-5-1(1) & 0.46 & 0.002\\
                           & AL2-W-5-1(2) & 0.655 & 0.003\\
                           & AL2-W-5-1(no-CIP) & 0.544 & 0.002\\
        \hline
        \multirow{2}{*}{2} & AL2-W-5-2(1) & 0.594 & 0.002\\
                           & AL2-W-5-2(2) & 0.577 & 0.002\\
        \hline
    \end{tabular}
\caption{Thermal diffusivity of active learning generated pellets}
\label{Table_AL_pellet_TD}
\end{center}
\end{table}

The corrected thermal conductivity of the pellets is shown in Table \ref{Table_AL_pellet_TC}. TC could not be calculated for the iteration 2 compositions as the methods used to calculate specific heat and bulk density in this study do not easily apply to dual-phase pellets. However, the values of the iteration 1 compositions match closely with the predictions of the surrogate model. This is despite the fact that the crystallographic parameters of the genrerated compositions are not an exact match for the parameters suggested by the AL. This is a promising indication that an exact match is not necessary in the context of optimizing low thermal conductivity compositions. Minor deviations in the compositions will likely result in minor deviations in thermal conductivity.

\begin{table}[H]
\begin{center}
    \begin{tabular}{lllll}
        \hline
        Iteration          & Composition & Corrected Thermal & St. Dev & Predicted\\
                           &             & conductivity (W/mK) & & Thermal Conductivity\\
        \hline
        \multirow{2}{*}{1} & AL-W-3-2 & 2.03 & 0.003 & 2.009\\
                           & AL-W-5-2 & 1.903 & 0.005 & 2.009\\
        \hline
        \multirow{3}{*}{2} & AL2-W-5-1(1) & -- & -- & 2.092\\
                           & AL2-W-5-1(2) & -- & -- & 2.092\\
                           & AL2-W-5-1(no-CIP) & -- & -- & 2.092\\
        \hline
        \multirow{2}{*}{2} & AL2-W-5-2(1) & -- & -- & 2.092\\
                           & AL2-W-5-2(2) & -- & -- & 2.092\\
        \hline
    \end{tabular}
\caption{Thermal conductivity of active learning generated pellets, corrected for porosity}
\label{Table_AL_pellet_TC}
\end{center}
\end{table}

\subsection{Machine learning model evaluation}

\begin{figure}[H]
    \centering
    \includegraphics[width=1\textwidth]{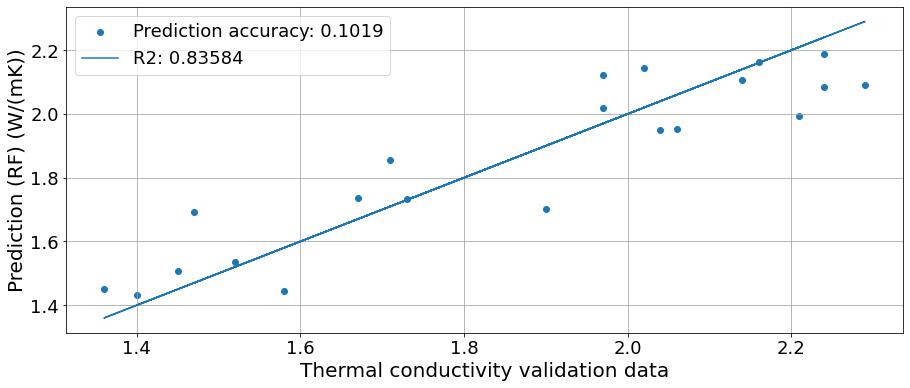}  
    \caption{Predicted thermal conductivity vs actual thermal conductivity of test data using surrogate model trained on the original dataset}
    \label{Img_pred_actual_TC_surrogate_model}
\end{figure}

The performance of the initial surrogate model (original dataset) demonstrates a $MAE$ sufficiently low for the range of TC and a high $R^2$ value as shown in Figure \ref{Img_pred_actual_TC_surrogate_model} and Table \ref{Table_surrogate_model_test_eval}. Here, dataset 2 refers to the original dataset plus the two compositions from the first iteration of the active learning process. Dataset 3 refers to dataset 2 plus the two compositions obtained from the second iteration of the active learning process. 

\begin{table}[H]
\begin{center}
  \begin{tabular}{cccccc}
    \hline
    \ {Iteration} &
      \multicolumn{2}{c}{MAE} &
       \multicolumn{2}{c}{$R^2$}
       & {Dataset}\\
    \hline
     & Mean & St. dev & Mean & St. dev\\
    \cmidrule(lr){2-2}\cmidrule(lr){3-3}\cmidrule(lr){4-4}\cmidrule(lr){5-5}\\
    0 & 0.102 & 0.069 & 0.836 & \textunderscore{} & original\\
    1 & 0.136 & 0.081 & 0.709 & \textunderscore{} & dataset 2\\
    2 & \textunderscore{} & \textunderscore{} & \textunderscore{} & \textunderscore{} & dataset 3\\
    \hline
  \end{tabular}
  \caption{Mean and standard deviation values of MAE and $R^2$ of surrogate model on test data on different stages of the active learning process}
  \label{Table_surrogate_model_test_eval}
\end{center}
\end{table}

\begin{figure}[H]
    \centering
    \includegraphics[width=1\textwidth]{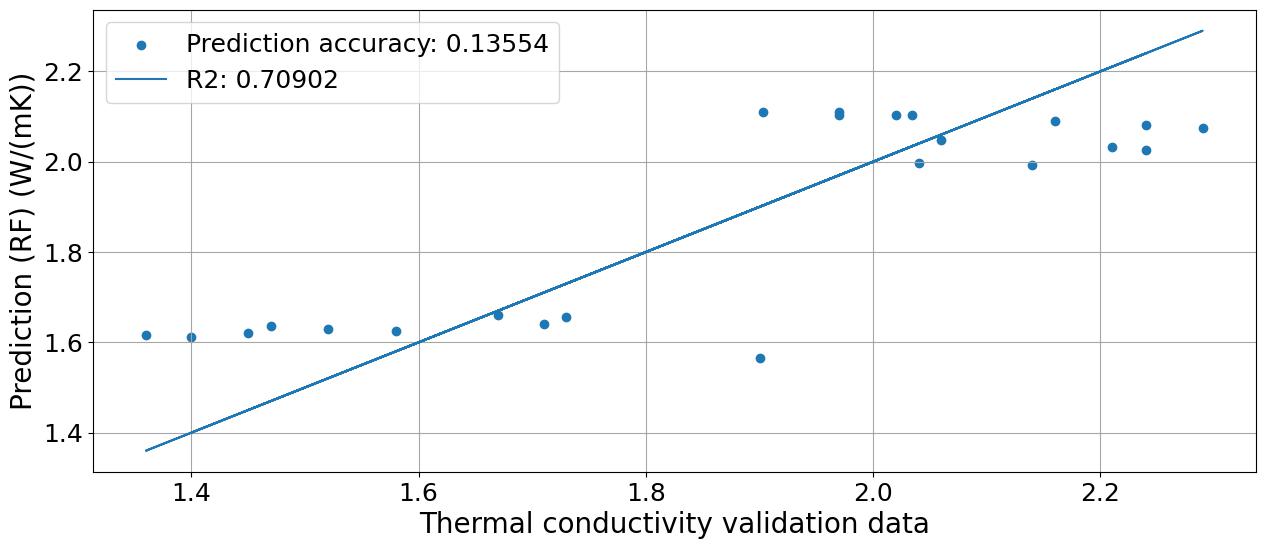}  
    \caption{Predicted thermal conductivity vs actual thermal conductivity of test data using surrogate model trained on dataset 2}
    \label{Img_pred_actual_TC_surrogate_model_Wright_AL1}
\end{figure}

\begin{figure}[H]
    \centering
    \includegraphics[width=1\textwidth]{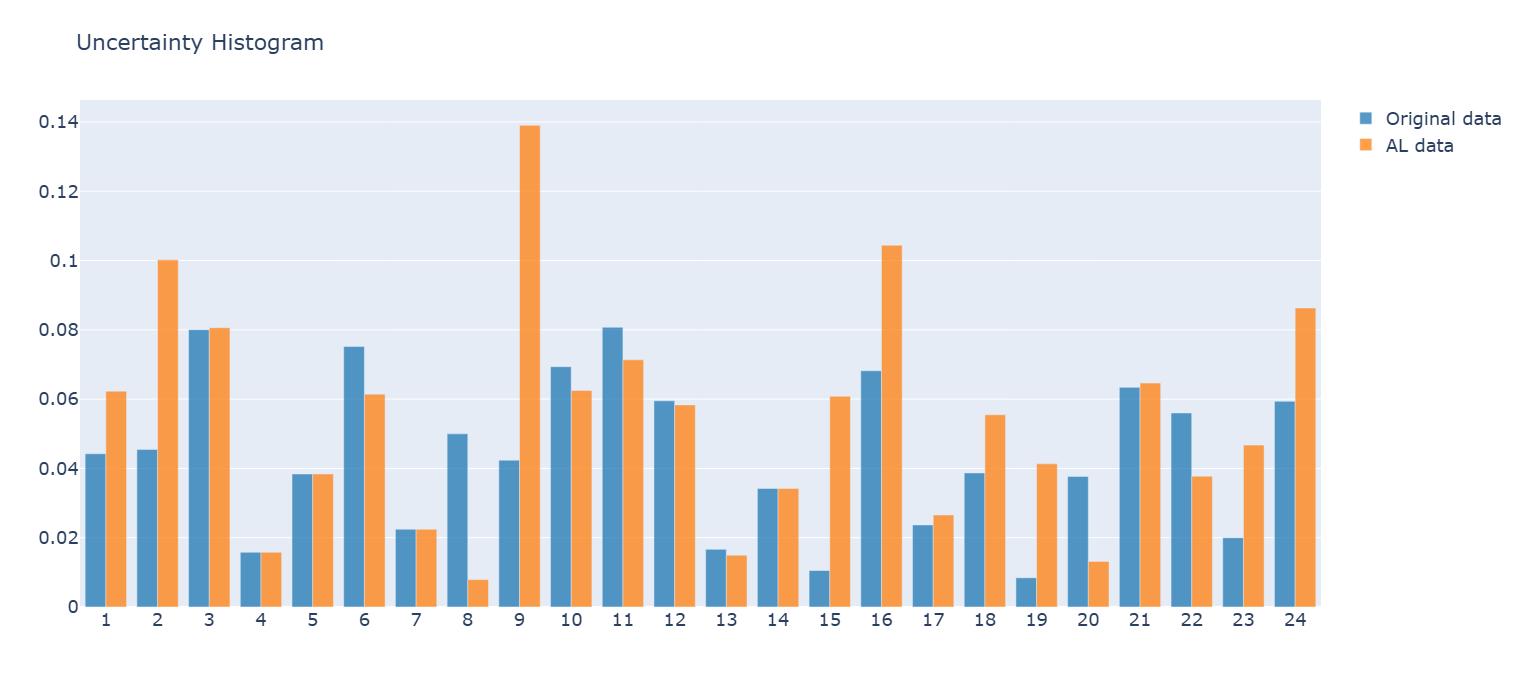}  
    \caption{Model uncertainty for each data-point before after adding active learning data-points}
    \label{Img_uncertainty_histogram}
\end{figure}

The addition of the active learning data from the first iteration appears to have lowered the performance. The predicted values appear to have split into two clusters as seen in Figure \ref{Img_pred_actual_TC_surrogate_model_Wright_AL1}. Effectively, the model appears to be associating certain features with TC values of roughly 1.6 $W/mK$ and the rest with 2.1 $W/mK$ values. Moreover, the uncertainty appears to have increased in parts of the dataset as seen in figure \ref{Img_uncertainty_histogram}. Due to the small size of the training data set, it is possible that the active learning does not have enough information to pinpoint a good area for exploration in the search space. The relatively high performance of the model, despite the sparsity of the dataset, could also be an indication of over-fitting. The compositions created in the second AL iteration could not be used to evaluate the performance of the second model due to being dual phase compositions (See section \ref{New_comp_characterisation}).

\subsection{Additional Data}

In dataset 2, additional data was introduced after the first iteration of the active learning process, from pellets that were developed in-house. The data consists of rare-earth zirconates with sets containing 2 and 3 elements in the $A$ cation position. Four sets were developed; lanthanum gadolinium, lanthanum yttrium, lanthanum ytterbium and lanthanum gadolinium yttrium, with each set containing various proportions of the rare-earth cations for a total of 19 extra data points. Dataset 3 is dataset 2 with outliers (high TC) removed. The training datasets are provided in the supplementary data.

\begin{figure}[H]
    \centering
    \includegraphics[width=1\textwidth]{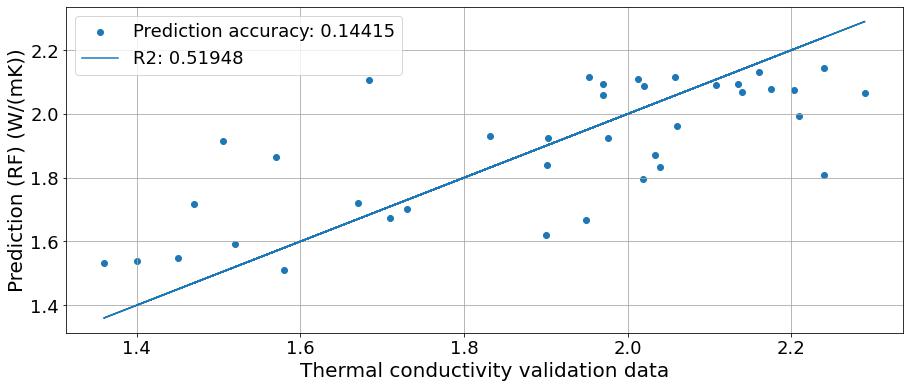}  
    \caption{Predicted thermal conductivity vs actual thermal conductivity of test data using surrogate model trained on dataset 3}
    \label{Img_pred_actual_TC_surrogate_model_Wright_AL1_RTZr_no_outliers}
\end{figure}

Figure \ref{Img_pred_actual_TC_surrogate_model_Wright_AL1_RTZr} shows the performance of the model retrained on dataset 2 with the additional experimental data. The $R^2$ is significantly lower after re-training. Removing the outliers ($Y_{2}Zr_{2}O{7}$, $La_{0.4}Y_{1.6}Zr_{2}O{7}$, $Gd_{2}Zr_{2}O{7}$ and $Yb_{2}Zr_{2}O{7}$) does not significantly change the performance (Figure \ref{Img_pred_actual_TC_surrogate_model_Wright_AL1_RTZr_no_outliers}). The new data points seem have more values with TC in the upper range of the distribution (2.0 to 2.2 $W/mK$). Furthermore, as all the in-house samples are zirconates, the $R_B$ and $M_B$ values are biased towards that of $Zr^{4+}$. That said, the model does not appear to be more accurate for zirconates vs any other composition (Table \ref{Table_top3_high_low_error_comp_wright_RTZr_AL1}).

\begin{figure}[H]
    \centering
    \includegraphics[width=1\textwidth]{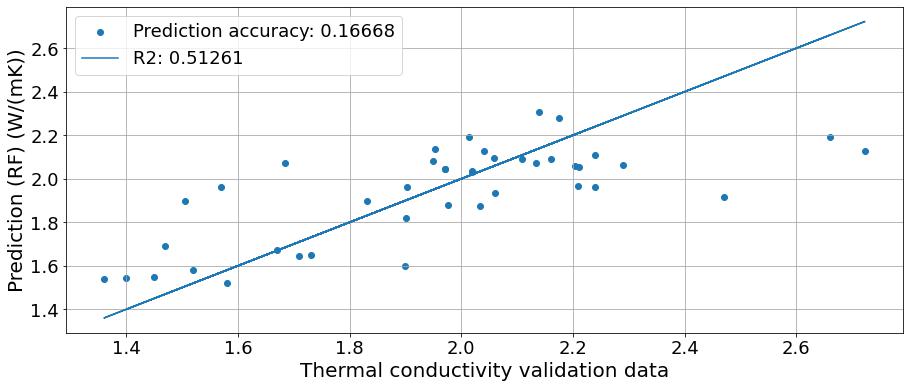}  
    \caption{Predicted thermal conductivity vs actual thermal conductivity of test data using surrogate model trained on dataset 2 plus additional data}
    \label{Img_pred_actual_TC_surrogate_model_Wright_AL1_RTZr}
\end{figure}

\begin{table}[H]
\begin{center}
    \begin{tabular}{lllll}
        \hline
        Error          & Composition & Actual TC & Predicted TC & MAE\\
        \hline
        \multirow{3}{*}{high} & $(La_{1/6}Gd_{2/3}Y_{1/6})_{2}Zr_{2}O_{7}$ &
                             1.684 & 2.107  & 0.423        \\
                           & $(Gd_{1/2}Eu_{1/2})_{2}(Zr_{1/3}Hf_{1/3}Sn_{1/3})_{2}O_{7}$   & 2.24 & 1.809 & 0.431        \\
                           & $(La_{1.2}Yb_{0.8})_{2}Zr_{2}O_{7})$ & 
                           1.506 & 1.915 & 0.409 \\
        \hline
        \multirow{3}{*}{low} & $(La_{1/3}Nd_{1/3}Pr_{1/3})_{2}(Zr_{1/2}Hf_{1/2})_{2}O_{7}$                       & 2.16 & 2.133 & 0.027 \\
                           & $(La_{1/3}Gd_{1/3}Y_{1/3})_{2}Zr_{2}O_{7}$     & 2.108 & 2.091 & 0.017 \\
                           & $(La_{0.333}Nd_{0.26}Gd_{0.15}Ho_{0.15}Yb_{0.111})_{2}Zr_{2}O_{7}$
                           & 1.903 & 1.23 & 0.02 \\
        \hline
    \end{tabular}
\caption{Actual TC, Predicted TC and MAE of the 3 compositions with the highest error and the 3 compositions with the lowest error from Figure \ref{Img_pred_actual_TC_surrogate_model_Wright_AL1_RTZr_no_outliers}}
\label{Table_top3_high_low_error_comp_wright_RTZr_AL1}
\end{center}
\end{table}

\subsection{Feature importances}

\begin{figure}[H]
    \centering
    \includegraphics[width=1\textwidth]{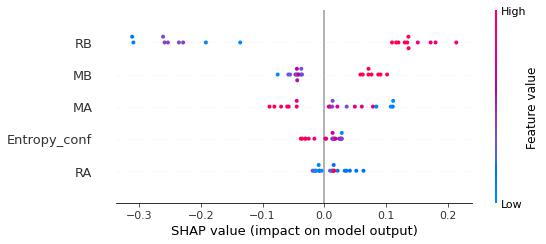}  
    \caption{Feature importances generated by \textit{SHAP} for the surrogate model trained on the original dataset}
    \label{Img_RF_kf7_feature_importance}
\end{figure}

\begin{figure}[H]
    \centering
    \includegraphics[width=1\textwidth]{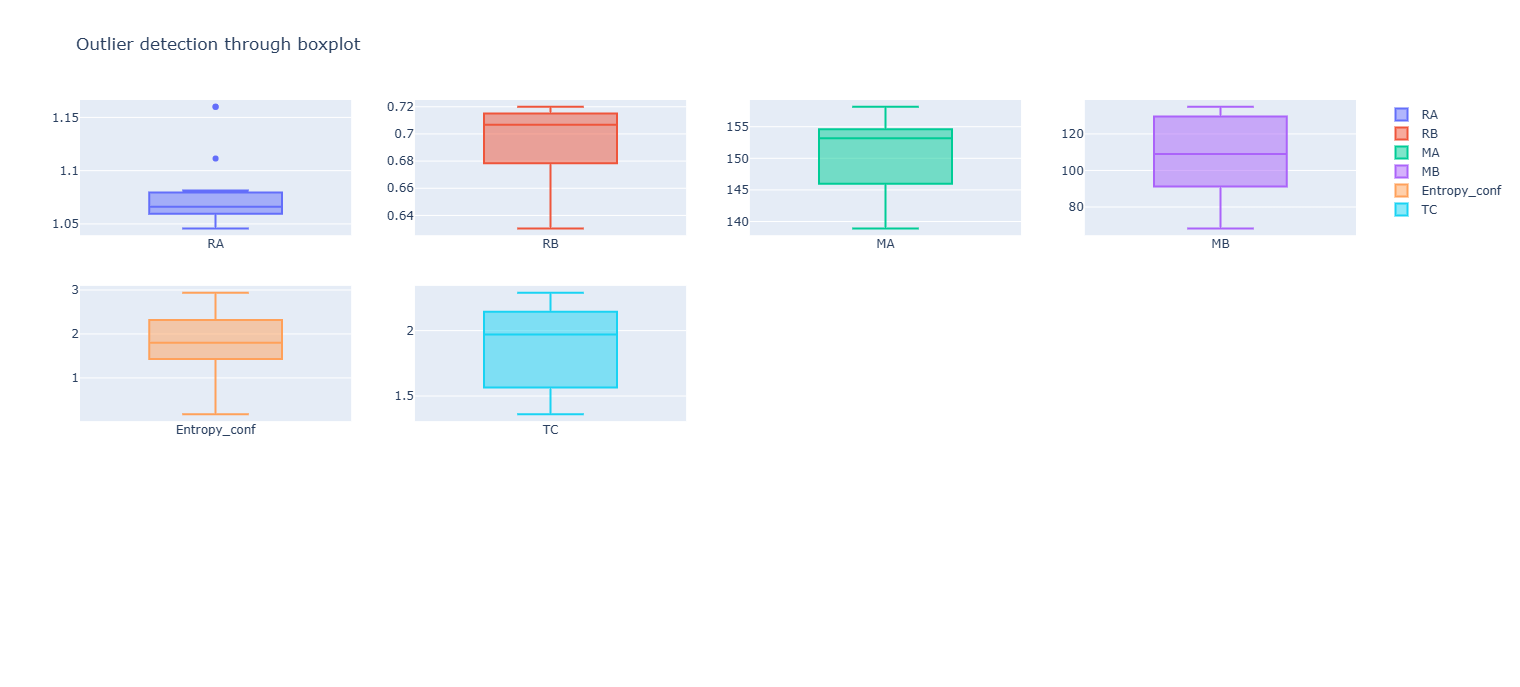}  
    \caption{Distribution of feature and thermal conductivity values within the literature training dataset}
    \label{Img_Wright_data_distribution}
\end{figure}

The \textit{SHAP} feature importances, for the surrogate model trained on the \textit{Wright et al}. dataset are shown in Figure \ref{Img_RF_kf7_feature_importance}. The plot indicates $R_{B}$ and $M_{B}$ rank highest and that higher values lead to higher thermal conductivity. Higher values of $M_{A}$ and configurational entropy seems to lead to lower thermal conductivity. $R_{A}$ ranks the lowest and it is difficult to ascertain the correlation between $R_{A}$ and thermal conductivity from the \textit{SHAP} plot due to the large difference between the outlier data and the rest of the dataset (Figure \ref{Img_Wright_data_distribution}).

\begin{figure}[H]
    \centering
    \includegraphics[width=1\textwidth]{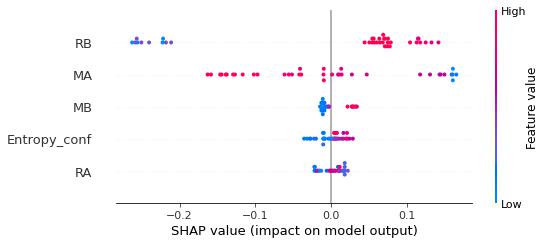}  
    \caption{Feature importances generated by \textit{SHAP} for the surrogate model trained on dataset 3}
    \label{Img_RF_Wright_RTZr_AL1_no_outlier_kf16_feature_importance}
\end{figure}

The model trained on dataset 3 shows a better distribution for $R_{A}$ (Figure \ref{Img_RF_Wright_RTZr_AL1_no_outlier_kf16_feature_importance}). The \textit{SHAP} values for this dataset shows most higher values of $R_{A}$ correlate with higher thermal conductivity, but there is also a mix of high and low values on each side. Going back to Figure \ref{Img_TC_LZ_GZ_series}, the relation between $R_A$ and TC appears to follow a second-order polynomial. It's difficult to say whether the \textit{SHAP} values displays the same relationship from Figure \ref{Img_RF_kf7_feature_importance}, the values in Figure \ref{Img_RF_Wright_RTZr_AL1_no_outlier_kf16_feature_importance} appears to conform to this relationship. 

$R_{B}$ still ranks highest in this dataset and shows a similar trend to the previous \textit{SHAP} values.

\begin{figure}[H]
    \centering
    \includegraphics[width=1\textwidth]{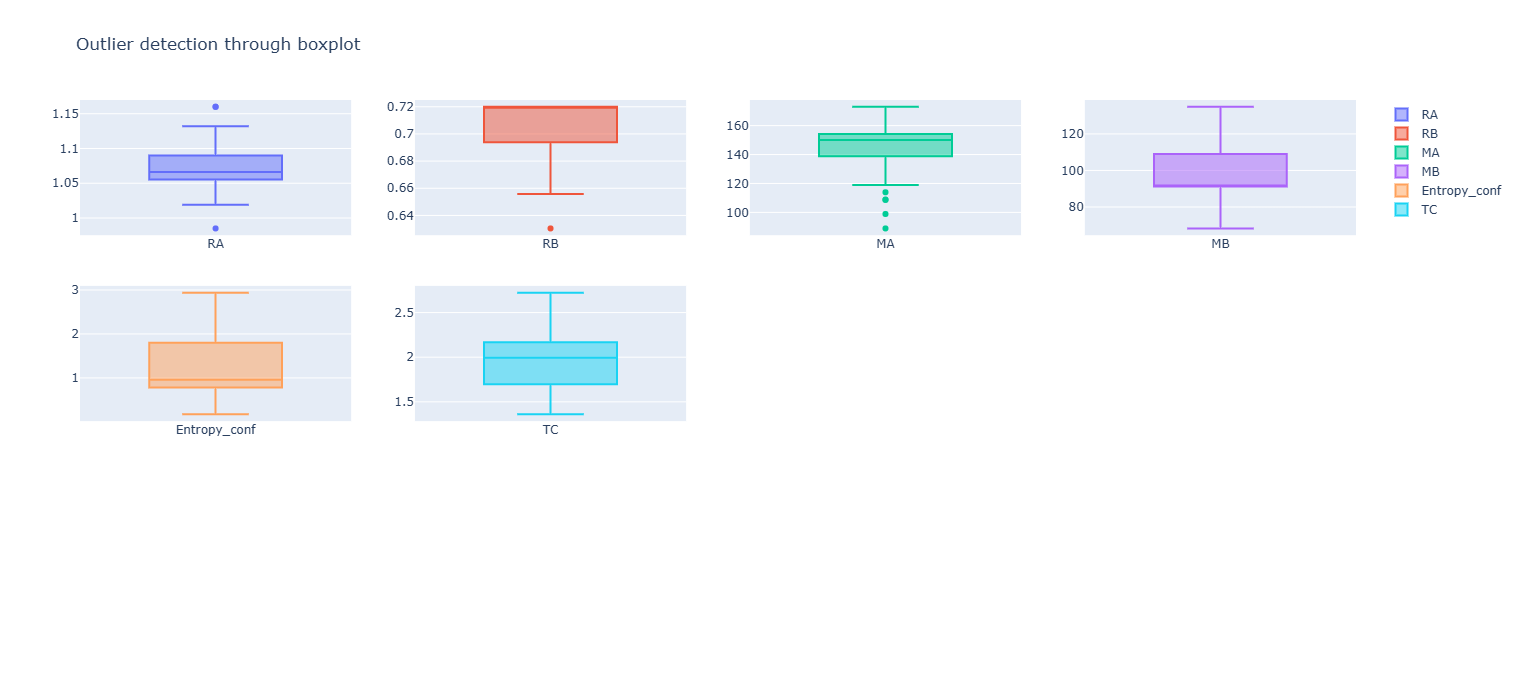}  
    \caption{Distribution of feature and thermal conductivity values within the combined literature and experimental training dataset}
    \label{Img_Wright_RTZr_data_distribution}
\end{figure}

\section{Conclusion}

In this study, active learning (AL) was used in conjunction with a Random Forest model to find low thermal conductivity (TC) high-entropy/multicomponent ceramic oxides (HEC and MCC) with pyrochlore structure. Four compositions were found, created and tested in the lab, over two iterations of the AL. The two compositions from the first iteration, $(La_{0.29}Nd_{0.36}Gd_{0.36})_{2}Zr_{2}O_{7}$ (AL-W-3-2) and $(La_{0.333}Nd_{0.26}Gd_{0.15}Ho_{0.15}Yb_{0.111})_{2}Zr_{2}O_{7}$ (AL-W-5-2), had measured thermal conductivities of 2.03  and 1.903 $W/mK$, respectively. The measured values matched closely with the predicted TC of the surrogate Random Forest model. However, the TC of these compositions fall between the median and third quartile values of TC present in the initial training dataset, with the minimum value in the training dataset being 1.36 $W/mK$. Given the sparsity of data with TC values between 1.8 and 2.0 $W/mK$, it is likely that the acquisition function prioritised exploration over exploitation. The compositions from the second iteration of AL, $(La_{0.43}Nd_{0.05}Gd_{0.05}Dy_{0.09}Yb_{0.38})_{2}Zr_{2}O_{7}$ (AL2-W-5-1) and $(La_{0.45}Nd_{0.05}Ho_{0.05}Dy_{1.0}\\Yb_{0.35})_{2}Zr_{2}O_{7}$ (AL2-W-5-2), turned out to be dual-phase, making it difficult to evaluate and compare the thermal conductivities of these compositions. The thermal diffusivity of the materials were significantly lower compared to the first set of AL compositions. However, this can be an effect of multiple phases and cannot be attributed to the AL process, as the surrogate model was trained to predict the TC of single-phase pyrochlores only. Further work is needed to draw any notable conclusions from this study. Constraints will need to be applied to the AL to ignore compositions that are potentially dual-phase, and more iterations of the AL will be required to determine if the system is capable of optimising compositions for low TC and/or efficiently improving the accuracy of the surrogate model.

\section*{Acknowledgements}
\noindent This work was supported by the Engineering and Physical Sciences Research Council (EPSRC) (grant number EP/V010093/1).

\section*{Data availability}

The data and code used for this study are available on a public repository on Github (\href{https://github.com/Roms89/Active_Learning_Driven_HEC_Rare-Earth_Oxide_discovery}{link here}).

\newpage
\bibliographystyle{ieeetr}
\bibliography{refs}

\begin{thebibliography}{10}

\bibitem{lagow_gas_turbine_temp_2016}
B.~W. Lagow, ``Materials selection in gas turbine engine design and the role of low thermal expansion materials,'' {\em JOM}, vol.~68, no.~11, p.~2770–2775, 2016.

\bibitem{mouritz_superalloys_2012}
A.~P. Mouritz, {\em 12 - Superalloys for gas turbine engines}, p.~251–267.
\newblock American Institute of Aeronautics and Astronautics, 2012.

\bibitem{Stecura1978EffectsOC}
S.~Stecura, ``Effects of compositional changes on the performance of a thermal barrier coating system.,'' Tech. Rep. NASA-TM-78976, NASA Lewis Research Center Cleveland, OH, United States, 1978.

\bibitem{cao_vassen_stoever_2004}
X.~Cao, R.~Vassen, and D.~Stoever, ``Ceramic materials for thermal barrier coatings,'' {\em Journal of the European Ceramic Society}, vol.~24, no.~1, p.~1–10, 2004.

\bibitem{Vasen2022-oq}
R.~Vaßen, E.~Bakan, D.~E. Mack, and O.~Guillon, ``A perspective on thermally sprayed thermal barrier coatings: Current status and trends,'' {\em Journal of Thermal Spray Technology}, vol.~31, no.~4, p.~685–698, 2022.

\bibitem{liu_HEC_2022}
D.~Liu, B.~Shi, L.~Geng, Y.~Wang, B.~Xu, and Y.~Chen, ``High-entropy rare-earth zirconate ceramics with low thermal conductivity for advanced thermal-barrier coatings,'' {\em Journal of Advanced Ceramics}, vol.~11, no.~6, p.~961–973, 2022.

\bibitem{darolia_2013}
R.~Darolia, ``Thermal barrier coatings technology: Critical review, progress update, remaining challenges and prospects,'' {\em International Materials Reviews}, vol.~58, no.~6, p.~315–348, 2013.

\bibitem{CMAS_multicomp_Rare_Earth_Li2023-om}
L.~Li, J.~Sun, C.~Li, W.~Lu, and Z.~Xu, ``{CMAS} resistance characteristics of multi-components rare earth phosphate materials at 1250 °c and 1350 °c,'' {\em Ceram. Int.}, vol.~49, pp.~39369--39383, Dec. 2023.

\bibitem{bakan_vaßen_2017}
E.~Bakan and R.~Vaßen, ``Ceramic top coats of plasma-sprayed thermal barrier coatings: Materials, processes, and properties,'' {\em Journal of Thermal Spray Technology}, vol.~26, no.~6, p.~992–1010, 2017.

\bibitem{mehboob_TBC_failure_mechanisms_2020}
G.~Mehboob, M.-J. Liu, T.~Xu, S.~Hussain, G.~Mehboob, and A.~Tahir, ``A review on failure mechanism of thermal barrier coatings and strategies to extend their lifetime,'' {\em Ceramics International}, vol.~46, no.~7, p.~8497–8521, 2020.

\bibitem{vassen_zirconates_TBC}
R.~Vassen, X.~Cao, F.~Tietz, D.~Basu, and D.~Stöver, ``Zirconates as new materials for thermal barrier coatings,'' {\em Journal of the American Ceramic Society}, vol.~83, no.~8, p.~2023–2028, 2004.

\bibitem{fergus_zirconia_pyrochlore_TBC_2014}
J.~W. Fergus, ``Zirconia and pyrochlore oxides for thermal barrier coatings in gas turbine engines,'' {\em Metallurgical and Materials Transactions E}, vol.~1, no.~2, p.~118–131, 2014.

\bibitem{wu_wei_low_TC_RE_Zirconates}
J.~Wu, X.~Wei, N.~P. Padture, P.~G. Klemens, M.~Gell, E.~García, P.~Miranzo, and M.~I. Osendi, ``Low-thermal-conductivity rare-earth zirconates for potential thermal-barrier-coating applications,'' {\em Journal of the American Ceramic Society}, vol.~85, no.~12, p.~3031–3035, 2004.

\bibitem{TC_zhu_meng_zhang_li_xu_reece_gao_2021}
J.~Zhu, X.~Meng, P.~Zhang, Z.~Li, J.~Xu, M.~J. Reece, and F.~Gao, ``Dual-phase rare-earth-zirconate high-entropy ceramics with glass-like thermal conductivity,'' {\em Journal of the European Ceramic Society}, vol.~41, no.~4, p.~2861–2869, 2021.

\bibitem{CWAN_LGZ}
C.~L. Wan, W.~Pan, Q.~Xu, Y.~X. Qin, J.~D. Wang, Z.~X. Qu, and M.~H. Fang, ``Effect of point defects on the thermal transport properties of ${({\mathrm{La}}_{x}{\mathrm{Gd}}_{(1\ensuremath{-}x)})}_{2}\mathrm{Zr}_{2}\mathrm{O}_{7}$: Experiment and theoretical model,'' {\em Physical Review B}, vol.~74, no.~14, 2006.

\bibitem{ren_LaYb_ReZr_HEC_2015}
X.~Ren, C.~Wan, M.~Zhao, J.~Yang, and W.~Pan, ``Mechanical and thermal properties of fine-grained quasi-eutectoid ($\mathrm{La}_{(1-x)}\mathrm{Yb}_x)_2\mathrm{Zr}_2\mathrm{O}_7$ ceramics,'' {\em Journal of the European Ceramic Society}, vol.~35, no.~11, p.~3145–3154, 2015.

\bibitem{liu_thermal_prprt_LGZ_2014}
S.~Liu, L.~Zhao, and K.~Jiang, ``Thermophysical properties of lanthanum-gadolinium zirconate and gadolinia-stabilized zirconia nanocomposite from in situ reaction,'' {\em Advanced Engineering Materials}, vol.~17, no.~3, p.~319–323, 2014.

\bibitem{tsai_HEA_2014}
M.-H. Tsai and J.-W. Yeh, ``High-entropy alloys: A critical review,'' {\em Materials Research Letters}, vol.~2, no.~3, p.~107–123, 2014.

\bibitem{liu_Materials_Discovery_2017}
Y.~Liu, T.~Zhao, W.~Ju, and S.~Shi, ``Materials discovery and design using machine learning,'' {\em Journal of Materiomics}, vol.~3, no.~3, p.~159–177, 2017.

\bibitem{Zhang2024_Pyrochlore_material_discovery_ML-dz}
Y.~Zhang, K.~Ren, W.~Y. Wang, X.~Gao, R.~Yuan, J.~Wang, Y.~Wang, H.~Song, X.~Liang, and J.~Li, ``Discovering the ultralow thermal conductive {A2B2O7-type} high-entropy oxides through the hybrid knowledge-assisted data-driven machine learning,'' {\em J. Mater. Sci. Technol.}, vol.~168, pp.~131--142, Jan. 2024.

\bibitem{Small_data_materials_science_Xu2023-re}
P.~Xu, X.~Ji, M.~Li, and W.~Lu, ``Small data machine learning in materials science,'' {\em Npj Comput. Mater.}, vol.~9, Mar. 2023.

\bibitem{Wright2019SizeDA}
A.~J. Wright, Q.~Wang, S.-T. Ko, K.~M. Chung, R.~Chen, and J.~Luo, ``Size disorder as a descriptor for predicting reduced thermal conductivity in medium- and high-entropy pyrochlore oxides,'' {\em Scripta Materialia}, 2019.

\bibitem{BayesOpt_mat_disc_Kusne2020-qd}
A.~G. Kusne, H.~Yu, C.~Wu, H.~Zhang, J.~Hattrick-Simpers, B.~DeCost, S.~Sarker, C.~Oses, C.~Toher, S.~Curtarolo, A.~V. Davydov, R.~Agarwal, L.~A. Bendersky, M.~Li, A.~Mehta, and I.~Takeuchi, ``On-the-fly closed-loop materials discovery via bayesian active learning,'' {\em Nat. Commun.}, vol.~11, p.~5966, Nov. 2020.

\bibitem{Pilania2021_Machine_learning_in_mat_sci}
G.~Pilania, ``Machine learning in materials science: From explainable predictions to autonomous design,'' {\em Comput. Mater. Sci.}, vol.~193, p.~110360, June 2021.

\bibitem{AL_HEA_Sulley2024-mf}
G.~A. Sulley, J.~Raush, M.~M. Montemore, and J.~Hamm, ``Accelerating high-entropy alloy discovery: efficient exploration via active learning,'' {\em Scr. Mater.}, vol.~249, p.~116180, Aug. 2024.

\bibitem{lookman_active_learning_materials_discovery_2019}
T.~Lookman, P.~V. Balachandran, D.~Xue, and R.~Yuan, ``Active learning in materials science with emphasis on adaptive sampling using uncertainties for targeted design,'' {\em npj Computational Materials}, vol.~5, no.~1, 2019.

\bibitem{AL_mat_discStein2019-ck}
H.~S. Stein and J.~M. Gregoire, ``Progress and prospects for accelerating materials science with automated and autonomous workflows,'' {\em Chem. Sci.}, vol.~10, pp.~9640--9649, Nov. 2019.

\bibitem{HEC_AL_mat_discvry_Leverant2024-rc}
C.~J. Leverant and J.~A. Harvey, ``Accelerating the discovery of new, single phase high entropy ceramics via active learning,'' {\em Chem. Mater.}, Nov. 2024.

\bibitem{HEC_vs_CCC_Wright2020-fi}
A.~J. Wright, Q.~Wang, C.~Huang, A.~Nieto, R.~Chen, and J.~Luo, ``From high-entropy ceramics to compositionally-complex ceramics: A case study of fluorite oxides,'' {\em J. Eur. Ceram. Soc.}, vol.~40, pp.~2120--2129, May 2020.

\bibitem{Park2023_mat_disc_AL_multi_obj_bayes_opt}
T.~Park, E.~Kim, J.~Sun, M.~Kim, E.~Hong, and K.~Min, ``Rapid discovery of promising materials via active learning with multi-objective optimization,'' {\em Mater. Today Commun.}, vol.~37, p.~107245, Dec. 2023.

\bibitem{Wang2021-pre-train_GP_for_Bayes_opt}
Z.~Wang, G.~E. Dahl, K.~Swersky, C.~Lee, Z.~Nado, J.~Gilmer, J.~Snoek, and Z.~Ghahramani, ``Pre-trained gaussian processes for bayesian optimization,'' {\em arXiv [cs.LG]}, 2021.

\bibitem{Active_learning_spray_Memon2024-qr}
H.~Memon, E.~Gjerde, A.~Lynam, A.~Chowdhury, G.~De~Maere, G.~Figueredo, and T.~Hussain, ``Active learning-driven uncertainty reduction for in-flight particle characteristics of atmospheric plasma spraying of silicon,'' {\em Eng. Appl. Artif. Intell.}, vol.~128, p.~107465, Feb. 2024.

\bibitem{Material_disc_bayes_opt_Chitturi2024-ux}
S.~R. Chitturi, A.~Ramdas, Y.~Wu, B.~Rohr, S.~Ermon, J.~Dionne, F.~H.~d. Jornada, M.~Dunne, C.~Tassone, W.~Neiswanger, and D.~Ratner, ``Targeted materials discovery using bayesian algorithm execution,'' {\em Npj Comput. Mater.}, vol.~10, July 2024.

\bibitem{Jin2023_bayes_opt_materials_discovery_review-xg}
Y.~Jin and P.~V. Kumar, ``Bayesian optimisation for efficient material discovery: a mini review,'' {\em Nanoscale}, vol.~15, pp.~10975--10984, July 2023.

\bibitem{Di_Fiore2024-AL_and_bayes_opt}
F.~Di~Fiore, M.~Nardelli, and L.~Mainini, ``Active learning and bayesian optimization: A unified perspective to learn with a goal,'' {\em Arch. Comput. Methods Eng.}, vol.~31, pp.~2985--3013, July 2024.

\bibitem{Exploration_Exploitation_Bondu2010-ss}
A.~Bondu, V.~Lemaire, and M.~Boulle, ``Exploration vs. exploitation in active learning : A bayesian approach,'' in {\em The 2010 International Joint Conference on Neural Networks ({IJCNN})}, IEEE, July 2010.

\bibitem{Xue2016-qa_Accel_material_design_adaptive_design}
D.~Xue, P.~V. Balachandran, J.~Hogden, J.~Theiler, D.~Xue, and T.~Lookman, ``Accelerated search for materials with targeted properties by adaptive design,'' {\em Nat. Commun.}, vol.~7, p.~11241, Apr. 2016.

\bibitem{Wei2025-cp_discv_lead_free_alloy_multi_bayes_opt}
Q.~Wei, Y.~Wang, G.~Yang, T.~Li, S.~Yu, Z.~Dong, and T.-Y. Zhang, ``Discovering novel lead-free solder alloy by multi-objective bayesian active learning with experimental uncertainty,'' {\em Npj Comput. Mater.}, vol.~11, Jan. 2025.

\bibitem{Shoyeb_Raihan2024-qa}
A.~Shoyeb~Raihan, H.~Khosravi, S.~Das, and I.~Ahmed, ``Accelerating material discovery with a threshold-driven hybrid acquisition policy-based bayesian optimization,'' {\em Manuf. Lett.}, vol.~41, pp.~1300--1311, Oct. 2024.

\bibitem{Farache2022-active_learning_high_melt_temp_alloy}
D.~E. Farache, J.~C. Verduzco, Z.~D. McClure, S.~Desai, and A.~Strachan, ``Active learning and molecular dynamics simulations to find high melting temperature alloys,'' {\em Comput. Mater. Sci.}, vol.~209, p.~111386, June 2022.

\bibitem{Van_Hoof2021-ns_hyperparameter_optimiz_surrogate_models}
J.~van Hoof and J.~Vanschoren, ``Hyperboost: Hyperparameter optimization by gradient boosting surrogate models,'' 2021.

\bibitem{suresh_TC_LZ_GZ}
G.~Suresh, G.~Seenivasan, M.~Krishnaiah, and P.~Srirama~Murti, ``Investigation of the thermal conductivity of selected compounds of gadolinium and lanthanum,'' {\em Journal of Nuclear Materials}, vol.~249, no.~2-3, p.~259–261, 1997.

\bibitem{leitner_Neumann-Kopp_application_2010}
J.~Leitner, P.~Voňka, D.~Sedmidubský, and P.~Svoboda, ``Application of neumann–kopp rule for the estimation of heat capacity of mixed oxides,'' {\em Thermochimica Acta}, vol.~497, no.~1-2, p.~7–13, 2010.

\bibitem{yang_mechanical_thermal_properties_RE_pyrochlores_2018}
L.~Yang, C.~Zhu, Y.~Sheng, H.~Nian, Q.~Li, P.~Song, W.~Lu, J.~Yang, and B.~Liu, ``Investigation of mechanical and thermal properties of rare earth pyrochlore oxides by first‐principles calculations,'' {\em Journal of the American Ceramic Society}, 2018.

\bibitem{Amiya_Data_representation_methods_2025-pl}
A.~Chowdhury, A.~R. Romero, E.~Aguilar-Bejarano, H.~Memon, G.~Figueredo, and T.~Hussain, ``A methodological study on data representation for machine learning modelling of thermal conductivity of rare-earth oxides,'' 2025.

\bibitem{Dippo_config_entropy_2021}
O.~F. Dippo and K.~S. Vecchio, ``A universal configurational entropy metric for high-entropy materials,'' {\em Scripta Materialia}, vol.~201, p.~113974, 2021.

\bibitem{shap_Lundberg2017-rg}
S.~Lundberg and S.-I. Lee, ``A unified approach to interpreting model predictions,'' 2017.

\bibitem{Bayesopt_EI_Brochu2010-ho}
E.~Brochu, V.~M. Cora, and N.~de~Freitas, ``A tutorial on bayesian optimization of expensive cost functions, with application to active user modeling and hierarchical reinforcement learning,'' 2010.

\bibitem{Explainable_ML_Mat_Sci_Zhong2022-kw}
X.~Zhong, B.~Gallagher, S.~Liu, B.~Kailkhura, A.~Hiszpanski, and T.~Y.-J. Han, ``Explainable machine learning in materials science,'' {\em Npj Comput. Mater.}, vol.~8, Sept. 2022.

\bibitem{MA_GZ_sintering_citric_acid_2015}
L.~MA, W.~MA, X.~SUN, J.~LIU, L.~JI, and H.~SONG, ``Structure properties and sintering densification of gd2zr2o7 nanoparticles prepared via different acid combustion methods,'' {\em Journal of Rare Earths}, vol.~33, p.~195–201, Feb 2015.

\bibitem{kubaschewski_alcock_evans_1979}
O.~Kubaschewski, C.~B. Alcock, and E.~L. Evans, {\em Metallurgical Thermochemistry}.
\newblock Pergamon Press, 1979.

\bibitem{fuentes_pyrochlore_ra_over_rb_criteria_2018}
A.~F. Fuentes, S.~M. Montemayor, M.~Maczka, M.~Lang, R.~C. Ewing, and U.~Amador, ``A critical review of existing criteria for the prediction of pyrochlore formation and stability,'' {\em Inorganic Chemistry}, vol.~57, no.~19, p.~12093–12105, 2018.

\bibitem{Teng2021-mz_HEC_phase_formation}
Z.~Teng, Y.~Tan, S.~Zeng, Y.~Meng, C.~Chen, X.~Han, and H.~Zhang, ``Preparation and phase evolution of high-entropy oxides {A2B2O7} with multiple elements at a and {B} sites,'' {\em J. Eur. Ceram. Soc.}, vol.~41, pp.~3614--3620, June 2021.

\bibitem{Liu2023-ik_HEC_phase_formation}
Z.~Liu, C.~Wei, Y.~De, S.~Zhang, C.~Zhang, and X.~Li, ``Phase structure of high-entropy pyrochlore oxides: From powder synthesis to ceramic sintering,'' {\em J. Eur. Ceram. Soc.}, vol.~43, pp.~7613--7622, Dec. 2023.

\end{thebibliography}

\newpage

\end{document}